\documentclass[aip,jcp,groupedaddress,a4paper,12pt,reprint]{revtex4-1}
\usepackage{amsmath}
\usepackage{amsfonts}
\usepackage{amssymb}
\usepackage{graphicx}

\usepackage{color}

\usepackage[makeroom]{cancel}
\usepackage{dcolumn}

\newcolumntype{.}{D{.}{.}{-1}}

\begin{document}
\title{Critical analysis of fragment-orbital DFT schemes for the calculation of electronic coupling values}
\author{Christoph Schober}

\author{Karsten Reuter}
\author{Harald Oberhofer}
\email{harald.oberhofer@ch.tum.de}
\affiliation{Chair for Theoretical Chemistry, Technische Universit{\"a}t M{\"u}nchen, Lichtenbergstr. 4, D-85747 Garching, Germany}
\date{\today}

\begin{abstract}
  We present a critical analysis of the popular fragment-orbital density-functional theory (FO-DFT) scheme for the calculation of electronic coupling values. We discuss the characteristics of different possible formulations or ''flavors'' of the scheme which differ by the number of electrons in the calculation of the fragments and the construction of the Hamiltonian. In addition to two previously described variants based on neutral fragments, we present a third version taking a different route to the approximate diabatic state by explicitly considering charged fragments. In applying these FO-DFT flavors to the two molecular test sets HAB7 (electron transfer) and HAB11 (hole transfer) we find that our new scheme gives improved electronic couplings for HAB7 ($-6.2\,\%$ decrease in mean relative signed error) and greatly improved electronic couplings for HAB11 ($-15.3\,\%$ decrease in mean relative signed error). A systematic investigation of the influence of exact exchange on the electronic coupling values shows that the use of hybrid functionals in FO-DFT calculations improves the electronic couplings, giving values close to or even better than more sophisticated constrained DFT calculations. Comparing the accuracy and computational cost of each variant we devise simple rules to choose the best possible flavor depending on the task. For accuracy, our new scheme with charged-fragment calculations performs best, while numerically more efficient at reasonable accuracy is the variant with neutral fragments.
\end{abstract}

\maketitle

\section{Introduction}
Charge transfer processes involving electrons or electron-holes are an integral part of many ubiquitous reactions, ranging from biological systems,\cite{moser_nature_1992,porath_direct_2000,gamiz-hernandez_linear_2015} and heterogenous catalysis\cite{munoz_ramo_theoretical_2007} to organic electronics\cite{coropceanu_charge_2007,painelli_electron-transfer_2007,clarke_charge_2010,bredas_molecular_2009,forrest_path_2004,facchetti_semiconductors_2007,dong_25th_2013,gershenson_colloquium:_2006} and many more. The theoretical description of such processes is commonly based on a diabatic picture of charge-localised initial and final states.\cite{wu_extracting_2006,van_voorhis_diabatic_2010} In this picture, one of the key factors determining the transport efficiency is the electronic coupling $H_{ab}$, also known as diabatic coupling or transfer integral. For two diabatic states $a$ and $b$, it is defined as
\begin{align}
  H_{ab} = \langle\Psi_a|\hat{\mathcal{H}}|\Psi_b\rangle,
\end{align}
where $\hat{\mathcal{H}}$ denotes the Hamiltonian of the system. Regardless of the actual model for charge transfer employed, such as small polaron hopping\cite{nelson_modeling_2009,oberhofer_revisiting_2012,saeki_comprehensive_2012,cornil_exploring_2013}, (coherent) band\cite{bardeen_deformation_1950,musho_ab-initio_2015} or polaronic band transport\cite{sanchez-carrera_theoretical_2010} -- see reference~\citenum{troisi_prediction_2007} for a detailed review -- the coupling elements contain most of the microscopic information such as e.g.~the relative geometry of molecular frontier orbitals partaking in the charge transfer.
Although there are -- depending on the exact charge transport mechanism -- a number of other parameters influencing the process, electronic couplings often serve as a first descriptor for gauging the charge transfer efficiency, especially in the field of organic electronics.\cite{mcgarry_rubrene-based_2013,mas-torrent_correlation_2004,garcia-frutos_crystal_2009,mitsui_dinaphtho[12-b:21-d]chalcogenophenes:_2013}
An accurate estimate of $H_\text{ab}$ is thus of great importance for the theoretical study of electron or hole transfer reactions.

In literature there are many methods to calculate electronic coupling values, with various degrees of accuracy and numerical efficiency. While very accurate electronic couplings can be evaluated using high-level quantum chemical calculations,\cite{cave_generalization_1996,cave_calculation_1997} their computational cost presently limits the role of these methods to benchmark calculations and small systems. In order to treat realistic systems a number of mainly density-functional theory (DFT) based methods have been developed. Prominent examples here are constrained density-functional theory (cDFT),\cite{wu_direct_2005,wu_extracting_2006,oberhofer_electronic_2010} frozen-density embedding\cite{te_velde_chemistry_2001} and the fragment-orbital (FO) methods.\cite{senthilkumar_charge_2003,oberhofer_revisiting_2012} Applications of these schemes range from biological systems,\cite{senthilkumar_charge_2003,kubar_efficient_2008,grozema_mechanism_2008} metal-oxides\cite{blumberger_constrained_2013,kerisit_molecular_2007} to the broad field of organic electronics, with organic solar cells,\cite{oberhofer_revisiting_2012,gajdos_inapplicability_2013,*gajdos_correction_2014,cornil_exploring_2013} organic light emitting transistors\cite{tamura_theoretical_2013} and organic field effect transistors.\cite{weng_diazapentacene_2009,karl_charge_2003,yi_charge-transport_2012,shinamura_synthesis_2010,mcgarry_rubrene-based_2013,sanchez-carrera_theoretical_2010} 
Among the more efficient schemes the FO approximation is by far the most popular, usually based on DFT\cite{karl_charge_2003,yi_charge-transport_2012,shinamura_synthesis_2010,mcgarry_rubrene-based_2013,sanchez-carrera_theoretical_2010} or even semi-empirical methods.\cite{kubas_electronic_2014,davino_energetics_2013,kirkpatrick_approximate_2008,ruhle_microscopic_2011} 

In the fragment (molecular) orbital method charge-localised diabatic states are constructed from non-interacting fragment densities. The fragments here correspond to donor and acceptor of the charge transfer process and ideally are separate molecular entities, such as for example neighboring molecules in a molecular crystal. Due to this simple approach, the method is easily incorporated in many modern electronic structure codes. While neglecting interactions between donor and acceptor sites during the calculation of the reference densities is of course a huge simplification and not feasible for all systems, it has been shown to work remarkably well for many applications in the field of organic electronics,\cite{kubas_electronic_2014,kubas_electronic_2015,oberhofer_revisiting_2012} where charge transfer sites are typically weakly interacting organic molecules. The method's efficiency makes it possible to investigate systems with many hundreds or thousands of different electronic coupling values.\cite{oberhofer_revisiting_2012} 

Yet, due to the approximations involved, the FO-DFT method should not be considered as a single method, but rather a family of methods. The basic steps are always the calculation of the fragment electron densities and the subsequent construction of the Hamiltonian from the superposition of the fragment densities. One of us has previously shown that in the original formulation of the method the number of electrons contributing to the Hamiltonian is wrong and corrected this by setting the occupation of the highest-occupied molecular orbital (HOMO) to zero in a second calculation step.\cite{oberhofer_electronic_2010} This method became known as FO-DFT($2n-1$), in contrast to the original FODFT($2n$) implementation. The performance of these methods compared to cDFT and high-level \emph{ab initio} reference data for hole and electron transfer systems (the HAB11+HAB7 data sets, respectively) was investigated in two recent studies.\cite{kubas_electronic_2014,kubas_electronic_2015} In this work we present another approach to consider the correct charge state in FO-DFT based on the calculation of appropriately charged fragments. This constitutes a third variant of FO-DFT which takes a slightly different route to arrive at the final approximation of the charge-separated diabatic states. All three methods have been implemented in the FHI-aims\cite{blum_ab_2009} program, allowing us to rule out any influences of different implementations and technical settings such as the choice of the basis set and enabling us to focus solely on the methodological differences in the schemes. We present results of systematic DFT calculations for all molecules in the references data sets (HAB11+HAB7) for all flavors of FO-DFT. To address the question of the accuracy of the methods we also compare electronic coupling values using gradient-corrected (GGA) and hybrid DFT functionals and analyze the effect of the exact exchange fraction in the construction of the Hamiltonian. In addition, we implemented a simple embedding approach to test the common interpretation that the neglected polarization of the reference densities in FO-DFT is responsible for the underestimation of electronic couplings. By analyzing this rich set of data we gain thorough insight into the accuracy and the computational efficiency of the different flavors of FO-DFT, allowing us to present guiding principles to select the best method depending on the desired accuracy and efficiency, for both hole and electron transfer electronic couplings.

The present work is organized as follows: First we review the theoretical background of the FO-DFT method and its variants, where we also introduce our new version which is based on charged-fragment calculations. Thereafter, we present our results for all three FO-DFT versions for the HAB11 and HAB7 test sets for hole and electron transfer, respectively. We end with a critical discussion of the merits and downsides of the different FO-DFT approaches both with GGA and hybrid level DFT, as well as the influence of neglecting inter-molecular polarization effects.

\section{Fragment-orbital DFT}\label{sec:theory}
In order to assess the different FO-DFT schemes, it is first necessary to review the theoretical background of the method. Special emphasis has to be put on the approximations underlying FO-DFT, as these give rise to a number of different formulations of this method. 

The basic idea behind the FO-DFT method is to construct the charge localized diabatic states from the isolated donor and acceptor fragments. In the following we exemplify this by a hole transfer between a donor $D^+$ and an acceptor $A$, with $n-1$ and $n$ electrons, respectively. Although not discussed here explicitly, electron transfer reactions, $D^- + A \rightarrow D + A^-$, can be treated in an entirely analogous manner. Furthermore, for clarity we restrict our derivation to symmetric cases where $D$ and $A$ are identical, but again, the derivation for asymmetric sites follows the same pattern.

\subsection{Kohn-Sham determinants}
Starting from the initial and final diabatic state single determinant wave functions,
\begin{subequations}
\begin{align}
        \Psi_a &= \frac{1}{(2n-1)!}\det\left(\phi_a^1,\ldots,\phi_a^{2n-1}\right)\\
        \Psi_b &= \frac{1}{(2n-1)!}\det\left(\phi_b^1,\ldots,\phi_b^{2n-1}\right) \quad,
\end{align}
\end{subequations}
the first step is to approximate the wave functions by the Kohn-Sham determinants of the isolated fragments,
\begin{subequations}
\label{eq:fodft_states}
\begin{align}
\Psi_a &\approx \Psi_a^{D^+A} = |\hat{\mathcal{A}} \phi_D, \ldots, \phi_D^{n-1}, {\color{white}\phi_D^n,} \phi_A, \ldots, \phi_A^{n-1}, \phi_A^n\rangle\label{eq:emp_state}\\
\Psi_b &\approx \Psi_b^{DA^+} = |\hat{\mathcal{A}} \phi_D, \ldots, \phi_D^{n-1}, \phi_D^n, \phi_A, \ldots, \phi_A^{n-1}{\color{white}, \phi_A^n}\rangle\label{eq:occ_state} \quad,
\end{align}
\end{subequations}
with the antisymmetrising operator $\hat{\mathcal{A}}$.
In constructing initial and final wave function from identical orbitals $\phi^{\left\{1, \ldots, n-1\right\}}_{D,A}$ one also assumes that only the frontier orbitals of the fragments differ, while all other orbitals are unchanged. Since neither sets of Kohn-Sham orbitals are eigenstates of the combined system's ground state Hamiltonian, they are generally not orthogonal. Therefore, in order to more closely resemble the charge separated diabatic states, the two sets of orbitals are orthogonalized, e.g.~with L\"owdin's symmetric scheme.\cite{lowdin_nonorthogonality_1950}
Also, there are in general two distinct coupling elements, one for the forward and one for the backward reaction. Denoting by $\hat{\mathcal{H}}_a$ and $\hat{\mathcal{H}}_b$ the Hamiltonians which give rise to the respective diabatic states -- which in FO-DFT are simply constructed from the respective fragment densities -- these coupling elements are
\begin{subequations}
\begin{align}\label{eq:hab}
H_{\text{ab}} = \langle\Psi_a|\hat{\mathcal{H}}_b|\Psi_b\rangle
\end{align}
and
\begin{align}\label{eq:hba}
H_{\text{ba}} = \langle\Psi_b|\hat{\mathcal{H}}_a|\Psi_a\rangle \quad.
\end{align}
\end{subequations}
If donor and acceptor molecules are identical, $H_{\text{ab}} = H_{\text{ba}}$ due to the symmetry of the system. 

By applying the Slater-Condon rules (see appendix~\ref{sec:slater} for a detailed derivation), the Kohn-Sham-determinant wave functions in eqs.~\ref{eq:fodft_states} are simplified to
\begin{subequations}\label{eq:simplified_ds_all}
\begin{align}\label{eq:simplified_ds}
&\Psi_a \approx \Psi_a^{D^+A} = | \cancel{\hat{\mathcal{A}}\phi_D, \ldots, \phi_D^{n-1}}, {\color{white}\phi_D^n,} \cancel{\phi_A, \ldots, \phi_A^{n-1}}, \phi_A^n\rangle\\
&\Psi_b \approx \Psi_b^{DA^+} = | \cancel{\hat{\mathcal{A}}\phi_D, \ldots, \phi_D^{n-1}}, \phi_D^n, \cancel{\phi_A, \ldots, \phi_A^{n-1}}{\color{white}. \phi_A^n}\rangle \quad,
\end{align}%
\end{subequations}
This reduces the calculation of electronic coupling elements to an integral between Kohn-Sham-orbitals
\begin{align}\label{eq:fo_dft_hab}
H_{ab} = \langle\phi_A^n|\hat{h}_{b}|\phi_D^n\rangle \quad,
\end{align}
with the single-particle Kohn-Sham Hamiltonian $\hat{h}_{b}$.
Note, that now $H_{ab}$ is determined for a pair of Kohn-Sham-orbitals instead of the full Kohn-Sham wave function in eq.~\ref{eq:hab}. We want to stress that this is a direct consequence of the previous approximation (eqs.~\ref{eq:emp_state} and \ref{eq:occ_state}) to the diabatic states and therefore exact within this representation. 

\subsection{Summary of approximations}\label{sec:approximations}
Up to this point we derived the general equations behind the FO idea. Before discussing the different flavors in which this method can be implemented we summarize the approximations made so far. 

\textbf{I:} The charge localized diabatic wave functions are approximated by Kohn-Sham determinants. 

\textbf{II} The Kohn-Sham determinants are constructed using reference densities calculated for isolated fragments, neglecting any interactions between the fragments that would result in a change of the self-consistent electron density and assuming that only the frontier orbitals change with the transferring charge.

\textbf{III:} The resulting charge localized Kohn-Sham wave functions are orthogonalized to resemble the diabatic state.

Electronic couplings according to eq.~\ref{eq:fo_dft_hab} depend on the Kohn-Sham orbitals $\phi_D^{n}$ and $\phi_A^{n}$, the HOMOs of the neutral hole transfer sites. In addition, one needs to determine the Hamiltonian $\hat{h}_b$ for the diabatic state $D^+A$, that is, for a wave function constructed from the Kohn-Sham orbitals $\phi_D,\ldots,\phi_D^{n-1}$ and $\phi_A,\ldots,\phi_A^{n}$. As within the FO-DFT approach either the neutral fragment \emph{or} the charged diabatic state is available, this necessitates the introduction of an additional approximation. Either the Hamiltonian $\hat{h}_b$ or the (frontier) orbitals $\phi_D^n$ and $\phi_A^n$ need to be approximated. This fundamental choice gives rise to the different formulations or ''flavors'' of FO-DFT known in literature. 

\subsection{Flavors of FO-DFT}\label{sec:flavours}
To allow a clear distinction between the different FO schemes we introduce the notation $\mathcal{H}^m@D^p A^q$, where $m$ is the number of electrons used to construct the Hamiltonian and $p, q \in \{+, -\}$ are the charges of donor and acceptor fragment, respectively. The original version of FO-DFT\cite{senthilkumar_charge_2003} using uncharged-fragment calculations and the subsequent $2n$-Hamiltonian is thus denoted as $\mathcal{H}^{2n}@DA$ in this notation. The number of electrons in the construction of the Hamiltonian is always given with respect to the number of electrons $n$ of a single neutral fragment.

\subsubsection{$\mathcal{H}^{2n}@DA$}
The original implementation by \citet{senthilkumar_charge_2003} within the ADF framework\cite{te_velde_chemistry_2001} made use of the ability of ADF to use molecular orbitals as basis set in the subsequent dimer calculation. The Hamiltonian $\hat{h}_b$ constructed this way is based on neutral fragments and $2n$ electrons. This means that while the orbitals $\phi_D$ and $\phi_A$ for the calculation of $H_{\text{ab}}$ are correct (within the approximation), the Hamiltonian is not.\cite{kubas_electronic_2014} This approach is very simple to implement in most electronic structure codes and therefore widely used.\cite{mitsui_dinaphtho[12-b:21-d]chalcogenophenes:_2013,shinamura_synthesis_2010,mitsui_naphtho[21-b:65-b]difuran:_2012,osaka_naphthodithiophenes_2013,wei_theoretical_2014,oberhofer_electronic_2010}

\subsubsection{$\mathcal{H}^{2n-1}@DA / \mathcal{H}^{2n+1}@D^-A^-$}
In our previous implementation\cite{oberhofer_revisiting_2012} of the method in the CPMD program\cite{cpmd} neutral fragments are used as well, but in the subsequent construction of the Hamiltonian the occupation number of the $\phi_D^n$ orbital is set to zero. This mimicks the correct charge in the diabatic states $\Psi_{a,b}$ and therefore the resulting Hamiltonian is based on the correct number of $2n-1$ electrons. The electronic coupling is then calculated between the LUMO of the donor and the HOMO of the acceptor. 

\subsubsection{$\mathcal{H}^{2n-1}@D^+A / \mathcal{H}^{2n+1}@D^-A$}
In addition to the two hitherto proposed methods, there is also a third possibility to construct reference states and Hamiltonian within FO-DFT. Instead of adjusting the occupation numbers in the Kohn-Sham orbitals in the second calculation step to get the correct number of electrons in the Hamiltonian, we here explicitly perform an SCF cycle on charged fragments. This has the advantage that the constructed Hamiltonian more closely resembles the correct Hamiltonian $\hat{\mathcal{H}}_b$ (as our constructed diabatic state is $\Psi^{D^+A}$, in contrast to $\Psi^{DA}$ in the other approaches). While the Hamiltonian is now correct, the frontier orbitals in eq.~\ref{eq:fo_dft_hab} differ
\begin{align}
H'_{ab} = \langle\phi_A^n|\hat{\mathcal{H}}|\phi_D^{n+}\rangle \quad,
\end{align}
with $\phi_D^{n+}$ being the LUMO of the calculated charged donor fragment $D^+$. As a consequence, the HOMO of the neutral donor $D$ is approximated by the LUMO of the charged fragment, $D^+$. While these can differ, we will show in section~\ref{sec:res_hab11hab7} later that for typical organic charge-transfer systems this approximation is less severe than approximating the correct diabatic Hamiltonian with neutral fragments. 

\subsection{Polarisation effects between fragments}\label{sec:embedding}
Due to the separation of the complete system into donor and acceptor fragments any polarization of the electron density of one fragment by the other, as it would occur for example in a cDFT calculation, is neglected in FO-DFT. Only in the final, non self-consistent calculation step, where the Hamiltonian is constructed based on the combined reference densities, the full exchange, correlation and electrostatic interactions are incorporated. To investigate the influence this has on the electronic couplings we implement a simple embedding approach. We here focus on the polarization due to Coulomb interactions between the fragments, as this should be the dominant effect in charged systems.
For this, we add to the wave function optimization of one fragment the full local potential ($V_{\text{local}}$) of the respective other fragment as an external potential. To achieve this $V^{\text{D+}}_{\text{local}} = V^{\text{D+}}_{\text{core}} + V^{\text{D+}}_{\text{elec}}$,
the local potential of the donor $D^+$, is added to the potential of the acceptor calculation $A$:
\begin{align}
  V^{\text{A}^{\delta+}}_{\text{local}} = V^{\text{A}}_{\text{core}} + V^{\text{A}}_{\text{elec}} + V^{\text{D+}}_{\text{local}}
\end{align}
This polarized variant of FO-DFT will be labeled $\delta^+$-FO-DFT. Here, only the neutral fragment is embedded into the local potential of the charged reaction partner, while in principle also the charged fragment is influenced through the presence of the neutral one. Yet, concentrating on electrostatic effects, such influences are minute and are therefore omitted.

\section{Computational details}
All calculations were performed with the FHI-aims package,\cite{blum_ab_2009,zhang_numeric_2013} where we implemented all three variants of FO-DFT and the embedded FO-DFT version. Electronic wave functions were expanded in a tier-2 numeric atomic orbital basis and tight integration grids, if not indicated otherwise. All dimer geometries were taken from the supporting information of the HAB7\cite{kubas_electronic_2015} and the HAB11\cite{kubas_electronic_2014} papers, respectively, and were used without further optimization, in order to pre-empt possible structural influences on the couplings. Total energies and coupling elements are calculated using the generalized gradient functional as proposed by Perdew, Burke and Ernzerhof (PBE)\cite{perdew_generalized_1996} and the Becke exchange functional in combination with the correlation functional by Lee, Yang, and Parr (BLYP).\cite{becke_density-functional_1988,lee_development_1988} In addition, the modified BLYP functional with a mixture of Hartree-Fock exact exchange, B3LYP\cite{becke_densityfunctional_1993,vosko_accurate_1980} (using the RPA version of the Vosk--Wilk--Nusair local density approximation), the PBE0 functional by Adamo and Barone\cite{adamo99} and the HSE06 functional by Heyd, Scuseria and Ernzerhof\cite{heyd03,heyd06,krukau06} were tested. 

Orthogonalization of the combined reference wave functions was achieved using the symmetric orthogonalization scheme by L\"owdin.\cite{lowdin_nonorthogonality_1950} As already noted in earlier work,\cite{kubas_electronic_2014,oberhofer_revisiting_2012,kubas_electronic_2015} special care has to be taken when calculating electronic couplings for degenerate states. 

To characterize the dependence of the electronic coupling on the donor-acceptor center to center distance $d$, the exponential decay $\beta$,
\begin{align}
  H_{ab} = A\cdot\exp\left(-\beta d/2\right),
\end{align}
is calculated for each system. To avoid an overvaluation of the small couplings due to the exponential function, we employed a linear regression on the logarithmized equation.

To exploit the symmetry between fragments within FO-DFT in FHI-aims we implemented the rotation of wave functions using Wigner~D matrices.\cite{wigner_eugen_gruppentheorie_1931} This allows us to re-use a once calculated density for all symmetrically identical fragments, thereby greatly reducing the computational cost of determining matrix elements for many different geometries (e.g.~in amorphous phases or organic crystals). While different schemes for real spherical harmonics are available (see for example \citet{lessig_efficient_2012,aubert_alternative_2013,blanco_evaluation_1997}), we construct our rotation matrices for different $l$ starting from the complex Wigner~D matrices via a transformation matrix $\mathbf{C}^l$. Using this approach it is easy to account for different sign conventions in the real $Y_{lm}$s employed by FHI-aims. The real rotation matrix $\Delta^l(R)$ is then obtained via
\begin{align}
        \Delta^l(R) = \left(\mathbf{C}^l\right)^*\mathbf{D}^l(R)\left(\mathbf{C}^l\right)^t,
\end{align}
and is used to obtain the rotated coefficients $\mathbf{c'}$ for each $l, m$ for each basis function in the system. This rotated Kohn-Sham wave function is then used for the new fragment geometry. The details of this method in the context of the FHI-aims code are further described in appendix~\ref{sec:localized}.

\section{Results}

\subsection{Electron/hole couplings (HAB7 and HAB11)}\label{sec:res_hab11hab7}
\begin{figure}[ht]
		\centering
		\includegraphics[scale=0.58,clip=true]{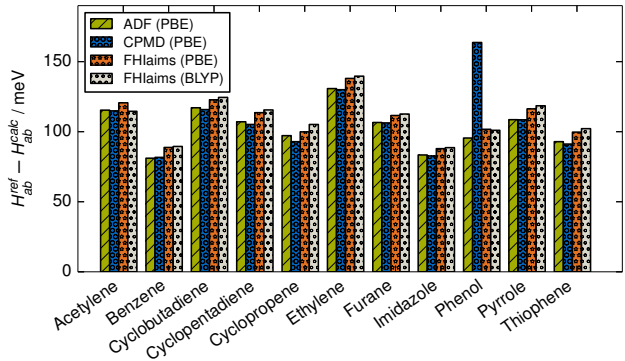}
    \caption{Comparison of electronic coupling values ($H_{ab}$ / meV) of the HAB11 test set computed with different electronic structure codes (FHI-aims, CPMD, ADF) against reference values published in Ref. \onlinecite{kubas_electronic_2014}. The large deviation in the value for phenol when calculated with the CPMD program has been noted before.\cite{kubas_electronic_2014} All couplings shown are for the $H^{2n}@DA$ variant of FO-DFT.}
		\label{fig:hab11_compare}
\end{figure}

\begin{figure}[ht]
		\centering
		\includegraphics[scale=0.58,clip=true]{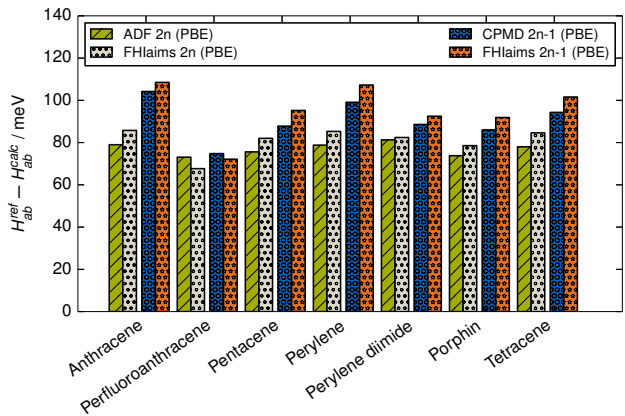}
    \caption{Comparison of electronic coupling values ($H_{ab}$ / meV) of the HAB7 test set computed with different electronic structure codes (FHI-aims, CPMD, ADF) against reference values published in Ref. \onlinecite{kubas_electronic_2015}. For each system the results for the $H^{2n}@DA$ and the $H^{2n+1}@D^-A^-$ flavor of FO-DFT are shown.} 
		\label{fig:hab7_compare}
\end{figure}

In order to rule out errors in our implementation and fluctuations due to differing basis sets and integration grids, we first compare the electronic coupling values for both test sets with their respective published values computed with the ADF and CPMD programs.\cite{kubas_electronic_2014,kubas_electronic_2015} For the HAB11 data set this is shown in Fig.~\ref{fig:hab11_compare}, while the numeric values are compiled in Table~S1 in the supporting information (SI).\cite{si} The comparison for the HAB7 electron transfer data set is shown in Fig.~\ref{fig:hab7_compare}, with the numeric results in Table~S2 of the SI.\cite{si} Although computed with three different electronic structure codes (using plane waves, Slater-type orbitals and numeric atomic orbitals as basis sets), the coupling values differ by less than 2\,\% -- viz.~$(1.91\pm 1.0)\,\%$ and $(1.30\pm 1.4)\,\%$ -- for the HAB11 and HAB7 set, respectively. The values for phenol calculated with CPMD have been omitted for the HAB11 set due to their known inaccuracy.\cite{kubas_electronic_2014}

\begin{figure}[ht]
		\centering
		\includegraphics[scale=0.58,clip=true]{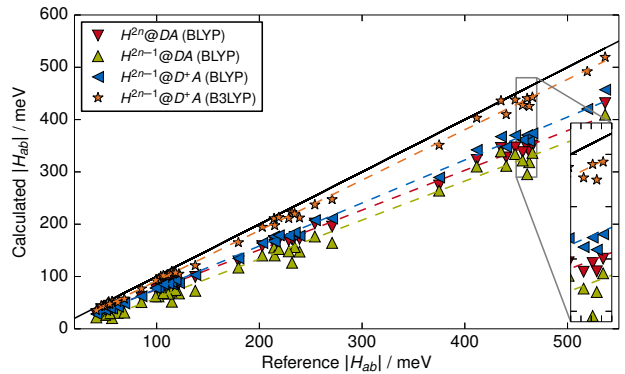}
    \caption{Correlation of electronic coupling values ($H_{ab}$ / meV) for the HAB11 data set with their \emph{ab initio} reference values for the different FO-DFT flavors at the GGA (BLYP) level and for the new flavor with charged-fragments additionally at the hybrid (B3LYP) level of theory.}
		\label{fig:hab11_correlation}
\end{figure}

Having established the consistency of our FHI-aims implementation with previous implementations in other DFT codes, we now assess the performance of the different flavors of FO-DFT presented in Sec.~\ref{sec:flavours}. As a reference, we compare our results with the high-quality ({\em ab initio}) benchmark data obtained by Kubas and co-workers.\cite{kubas_electronic_2014,kubas_electronic_2015} 
The correlation of our calculated electronic couplings for the HAB11 hole transfer database with the reference values is shown in Fig.~\ref{fig:hab11_correlation}, while all numerical values are compiled in Table~S3 in the SI.\cite{si} Consistent with previous results,\cite{kubas_electronic_2014} we find that all flavors of FO-DFT using GGA functionals underestimate the electronic couplings. The mean relative signed errors (MRSE) for this model lie between $-37.7\,\%$ ($H^{2n-1}@DA$) and $-22.4\,\%$ ($H^{2n-1}@D^+A$) for the BLYP functional, with equivalent findings for the PBE functional. A comprehensive overview of the methods' accuracy is given in Table~\ref{tab:stats_hab11}. Note that the new flavor based on charged-fragment calculations yields a significant improvement ($-22.4\,\%$ vs $-37.7\,\%$ at BLYP level of theory) over the uncharged variants of FO-DFT. This effect consistently occurs for all 11 systems of the test set. In addition, we also determined the decay of $H_\text{ab}$ with increasing donor-acceptor separation. Results included in Table \ref{tab:stats_hab11} show that the $H^{2n-1}@D^+A$ scheme also yields an improved decay constant $\beta$ compared to the other variants and the {\em ab initio} reference value.

\begin{figure}[ht]
		\centering
		\includegraphics[scale=0.58,clip=true]{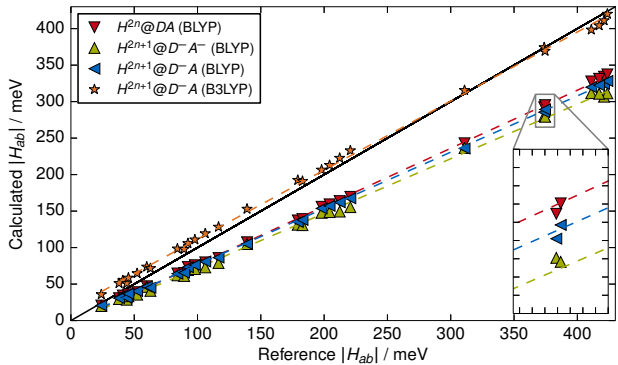}
    \caption{Correlation of electronic coupling values ($H_{ab}$ / meV) for the HAB7 data set with their \emph{ab initio} reference values for the different FO-DFT flavors at the GGA (BLYP) level and for the new flavor with charged-fragments additionally at the hybrid (B3LYP) level of theory.}
		\label{fig:hab7_correlation}
\end{figure}

The accuracy of all three methods with respect to electron transfer was determined by means of the HAB7 test set, again referencing to high level \emph{ab initio} results. The correlation of the different flavors and functionals with the reference values is summarized in Fig.~\ref{fig:hab7_correlation} with the numerical results compiled in Table~S4 in the SI.\cite{si} As was the case for the HAB11 test set, the new flavor based on charged fragments performs very well, with an MRSE of $-22.9\,\%$, compared to $-22.4\,\%$ and $-27.1\,\%$ for the $H^{2n}@DA$ and $H^{2n+1}@D^-A^-$ schemes, respectively. 

In order to gauge the influence of functional accuracy on calculated coupling values, we additionally computed all couplings with the hybrid functional B3LYP. Here, further improvements towards the reference values can be found for both test sets. In the case of hole transfer (HAB11), the MRSE for B3LYP is $-7.3\,\%$, which is comparable to the value obtained by the more sophisticated cDFT scheme with exact exchange ($13.8\,\%$).\cite{kubas_electronic_2014} To verify the generality of this effect, we also calculated electronic couplings for HAB11 with different hybrid functionals (PBE0, HSE06), obtaining the same overall trend. Exemplary values for furane are compiled in Table~\ref{tab:hybrids}. In the case of electron transfer (HAB7), the MRSE for B3LYP is $11.8\,\%$, even improving on earlier PBE0-cDFT results with an MRSE of $30.7\,\%$.\cite{kubas_electronic_2015} In contrast to the HAB11 data set, the B3LYP values slightly overestimate the electronic couplings. 

\begingroup
\squeezetable
\begin{table*}[ht]
\caption{Compilation of the mean unsigned error (MUE = $\left(\sum_n| H_{ab}^{\text{calc}} - H_{ab}^{\text{ref}}|\right) / n$ ), the mean relative signed error (MRSE = $\left(\sum_n\left(H_{ab}^{\text{calc}} - H_{ab}^{\text{ref}}\right) / H_{ab}^{\text{ref}}\right) /n $), the mean relative unsigned error (MRUE = $\left(\sum_n|H_{ab}^{\text{calc}} - H_{ab}^{\text{ref}}| / H_{ab}^{\text{ref}}\right) /n $) and the highest single absolute error (MAX = $\text{max}|H_{ab}^{\text{calc}} - H_{ab}^{\text{ref}}|$) for all electronic coupling values ($|H_{ab}|$) and for the distance decay constants ($\beta$) for the HAB11 data set.}
\begin{ruledtabular}
\begin{tabular}{lc......}
 \multicolumn{1}{l}{ } & \multicolumn{1}{c}{  } & \multicolumn{1}{l}{ $H^{2n}@DA$ } & \multicolumn{1}{l}{ $H^{2n-1}@DA$ } & \multicolumn{3}{l}{ $H^{2n-1}@D^+A$ } \\ \cline{5-8}
  &  &  &  &  &  &  & \\
 \multicolumn{1}{l}{ } & \multicolumn{1}{l}{  } &  \multicolumn{1}{l}{ BLYP } &  \multicolumn{1}{l}{ BLYP } &  \multicolumn{1}{l}{ BLYP } & \multicolumn{1}{l}{ B3LYP } & \multicolumn{1}{l}{ BLYP@B3LYP } & \multicolumn{1}{l}{ B3LYP@BLYP } \\
 $|H_{ab}|$ & MUE / meV & 51.7 & 69.6 & 43.7 & 12.4 & 45.7 & 6.6 \\
  & MRSE / \% &  -24.6 & -37.7 & -22.4 & -7.3 & -23.5 & 2.0 \\
  & MRUE / \% &  24.6 & 37.7 & 22.4 & 7.4 & 23.5 & 4.4 \\
  & MAX / meV &  139.7 & 165.2 & 103.9 & 37.1 & 107.9 & 22.2 \\
  &  &  &  &  &  &  &  \\
 $\beta$ & MUE / 1/\AA & 0.06 & 0.42 & 0.12 & 0.10 & 0.14 & 0.10 \\
  & MRSE / \% & 0.2 & 14.9 & 4.3 & 3.2 & 4.8 & -3.1 \\
  & MRUE / \% & 2.2 & 14.9 & 4.3 & 3.5 & 4.8 & 3.4 \\
  & MAX / 1/\AA & 0.12 & 0.77 & 0.21 & 0.17 & 0.24 & 0.23 \\
\end{tabular}
\end{ruledtabular}
\label{tab:stats_hab11}
\end{table*}
\endgroup

\begingroup
\squeezetable
\begin{table*}[ht]
  \caption{Same as Table \ref{tab:stats_hab11} for the HAB7 data set.}
\begin{ruledtabular}
\begin{tabular}{lc......}
 \multicolumn{1}{l}{ } & \multicolumn{1}{c}{  } & \multicolumn{1}{l}{ $H^{2n}@DA$ } & \multicolumn{1}{l}{ $H^{2n-1}@DA$ } & \multicolumn{3}{l}{ $H^{2n-1}@D^+A$ } \\ \cline{5-8}
  &  &  &  &  &  &  & \\
 \multicolumn{1}{l}{ } & \multicolumn{1}{l}{  } &  \multicolumn{1}{l}{ BLYP } &  \multicolumn{1}{l}{ BLYP } &  \multicolumn{1}{l}{ BLYP } & \multicolumn{1}{l}{ B3LYP } & \multicolumn{1}{l}{ BLYP@B3LYP } & \multicolumn{1}{l}{ B3LYP@BLYP } \\
 $|H_{ab}|$ & MUE / meV & 39.4 & 47.8 & 41.6 & 10.4 & 43.1 & 9.6 \\
  & MRSE / \% & -22.4 & -27.1 & -22.9 & 11.8 & -23.7 & 11.0 \\
  & MRUE / \% & 22.4 & 27.1 & 22.9 & 12.6 & 23.7 & 11.8 \\
  & MAX / meV & 86.9 & 114.4 & 96.7 & 14.3 & 103.1 & 14.5 \\
  &  &  &  &  &  &  &  \\
 $\beta$ & MUE / 1/\AA & 0.09 & 0.10 & 0.09 & 0.34 & 0.08 & 0.33 \\
  & MRSE / \% & 1.3 & 1.6 & -0.7 & -11.4 & -0.1 & -11.1 \\
  & MRUE / \% & 3.3 & 3.3 & 3.1 & 11.4 & 2.8 & 11.1 \\
  & MAX / 1/\AA & 0.18 & 0.19 & 0.17 & 0.51 & 0.15 & 0.50 \\

\end{tabular}
\end{ruledtabular}
\label{tab:stats_hab7}
\end{table*}
\endgroup

\begin{table}[ht]
	\caption{Electronic coupling values for furane at different dimer distances calculated with the B3LYP, PBE0 and HSE06 functionals. GGA-BLYP values are shown for comparison. All values in meV.}
\begin{ruledtabular}
	\begin{tabular}{l....}
 \multicolumn{1}{l}{  } & \multicolumn{1}{l}{ BLYP } & \multicolumn{1}{l}{ B3LYP } & \multicolumn{1}{l}{ PBE0 } & \multicolumn{1}{l}{ HSE06 } \\
 $5.0\,\text{\AA}$ & 36.8 & 43.8 & 44.4 & 41.7 \\
 $4.5\,\text{\AA}$ & 79.1 & 94.1 & 95.1 & 89.9 \\
 $4.0\,\text{\AA}$ & 166.7 & 197.7 & 200.6 & 191.6 \\
 $3.5\,\text{\AA}$ & 347.7 & 409.9 & 420.2 & 406.9 \\
\end{tabular}
\end{ruledtabular}
\label{tab:hybrids}
\end{table}

\subsection{Hybrid FO-DFT on GGA densities}\label{sec:crossover}
To further investigate the influence of the functional on calculated FO-DFT couplings, specifically the source of the improvement we see in the hybrid-level calculations, we also determine $H_\text{ab}$ using a hybrid-GGA crossover approach. As discussed above, there are essentially two parts to an FO-DFT calculation: the construction of the reference diabatic states and densities, and the subsequent generation of the diabatic states' Hamiltonian. To disentangle the influence of exact exchange on both parts we determine the couplings using GGA-BLYP densities in the construction of a B3LYP Hamiltonian (B3LYP@BLYP) and vice versa (BLYP@B3LYP).

\begin{figure}[ht]
    \centering
    \includegraphics[scale=0.58,clip=true]{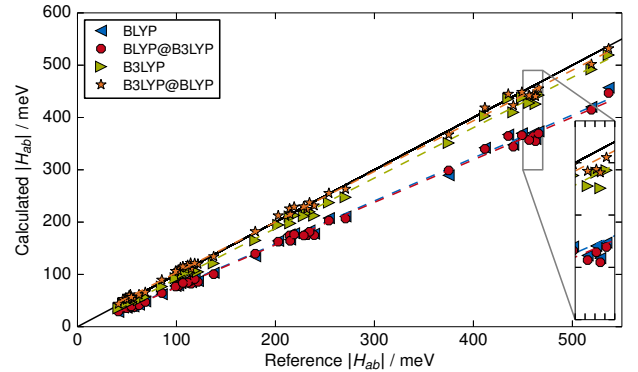}
    \caption{Correlation between calculated and reference electronic coupling values for the HAB11 data set. The BLYP and B3LYP values show couplings calculated with the same functional for the self-consistent fragment density and the dimer Hamiltonian, while BLYP@B3LYP and B3LYP@BLYP refer to the hybrid-GGA crossover schemes, see text.}
    \label{fig:gga_hybrid_correlation}
\end{figure}

\begin{figure}[ht]
    \centering
    \includegraphics[scale=0.58,clip=true]{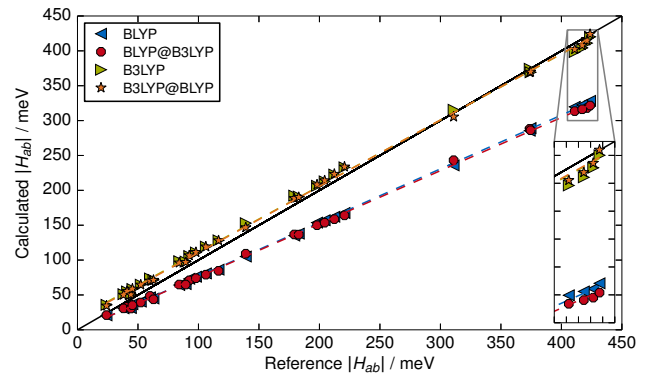}
    \caption{Correlation between calculated and reference electronic coupling values for the HAB7 data set. The BLYP and B3LYP values show couplings calculated with the same functional for the self-consistent fragment density and the dimer Hamiltonian, while BLYP@B3LYP and B3LYP@BLYP refer to the hybrid-GGA crossover schemes, see text.} 
    \label{fig:hab7_zombie}
\end{figure}

In our FO-DFT calculation using the hybrid-functional reference density (BLYP@B3LYP) we see lower couplings due to a reduced wave function overlap. This is expected as hybrid functionals generally yield a higher degree of localization of the electron density.\cite{cohen_insights_2008} Compared to the pure GGA results the values decrease on average by $1.39\,\%$ and $1.08\,\%$ for the HAB11 and HAB7 data set, respectively. The inverse procedure, with the Hamiltonian constructed at the B3LYP level, but based on the GGA-BLYP densities, on the other hand, yields largely improved electronic couplings, with an average increase of the couplings by $31.8\,\%$ and $43.8\,\%$ for HAB11 and HAB7, respectively, as compared to the pure GGA results. The resulting values are then very close to the pure hybrid results. This is shown in Fig.~\ref{fig:gga_hybrid_correlation} for the HAB11 data set and in Fig.~\ref{fig:hab7_zombie} for the HAB7 set. 

\subsection{Polarized FO-DFT ($\delta^+$-FO-DFT)}
\begin{figure}[ht]
 \centering
 \includegraphics[scale=0.4,clip=true]{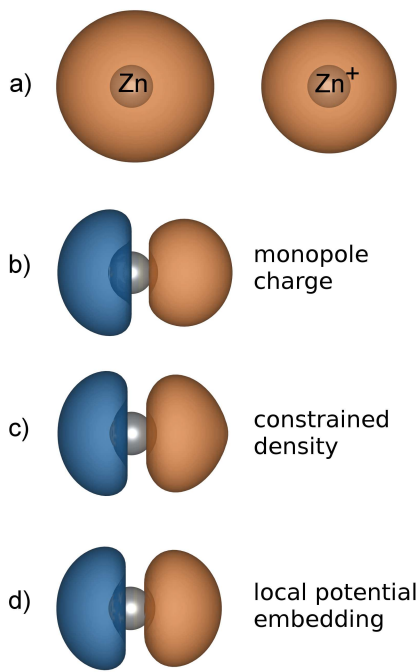}
 \caption{a) Total electron density for a Zn$_2^+$-dimer (distance $5\,\text{\AA}$) calculated with cDFT. b-d) Density difference between the free atom density for Zn and the calculated total density with monopole, cDFT and electrostatic potential embedding, respectively. For each method, the electron density distorts from the free atom density as a consequence of the nearby positive charge. The comparison reveals good agreement between the cDFT and the embedding result, while the distortion due to the monopole is slightly more pronounced towards the positive charge and carries the danger of electron spill-out.\cite{berger_embedded-cluster_2014}}
 \label{fig:electrostatic_scheme}
\end{figure}

As outlined above, a frequently suspected reason for the underestimation of couplings in FO-DFT is the neglect of polarization of the fragment densities. To estimate the effect this really has on the couplings we perform calculations with the local-potential embedding scheme outlined in section \ref{sec:embedding}. In Fig.~\ref{fig:electrostatic_scheme} we illustrate the fragment polarization through electrostatic potential embedding for the example of a Zinc dimer and compare it to other approaches. Once, the Zn$^+$ atom is replaced by a positive point charge at the position of the nucleus, and once the positive charge is constrained to the atom using cDFT. The effect of the prior pure electrostatic embedding on the density is very similar to the constrained DFT density, showing that it is in principle possible to approximately include polarization effects in the fragment calculation.

\begin{figure}[ht]
		\centering
		\includegraphics[scale=0.58,clip=true]{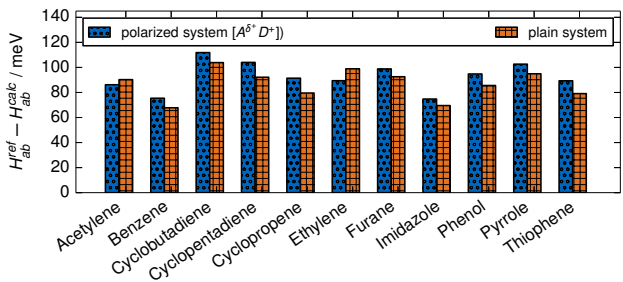}
    \caption{Comparison of electronic couplings ($H_{ab}$ / meV) calculated with the plain FO-DFT method and with polarized $\delta^+$-FODFT. All couplings are calculated with the BLYP functional and the $H^{2n-1}@D^+A$ variant of FO-DFT.}
		\label{fig:hab11_embedding}
\end{figure}

Applied to the HAB11 data set, we find that using polarized fragment densities has only a small influence on the coupling values. The average change in electronic couplings for all systems is only $1.65\,\%$, with the highest single change being $3.2\,\%$ (cf.~Fig. \ref{fig:hab11_embedding}). 

\section{Discussion}

\subsection{The effect of charged-fragment calculations}
Our new variation of the FO-DFT scheme, which is based on charged-fragment calculations (cf.~section~\ref{sec:theory}), yields much improved electronic couplings for the HAB11 hole transfer test set. In order to understand why, we again turn to the two steps of FO-DFT, 1) the approximation of the frontier orbitals of the diabatic charge transfer states and 2) the construction of the diabatic Hamiltonian. The $\{H^{2n}/H^{2n-1}\}@DA$ variants share the correct (neutral) frontier orbitals and both approximate the Hamiltonian to some extent. The $2n-1$ scheme used in the CPMD implementation allows for the correct number of electrons in the Hamiltonian, while the $2n$ scheme always uses the neutral dimer as reference system. Although electronic couplings obtained with the $2n$ scheme show better agreement with the reference, it has been shown that this is due to fortunate error compensation.\cite{kubas_electronic_2014}

The $H^{\text{2n-1}}@D^+A$ scheme, on the other hand, has the correct number of $2n-1$ electrons by construction -- due to the charged fragment calculation, but approximates the neutral frontier orbital by the LUMO, i.e.~the unoccupied minority spin orbital of a singly occupied molecular orbital. Since the electronic coupling is then calculated between the unoccupied minority spin orbital $\chi^{n/2}$ of (the charged) fragment 1 and the occupied minority spin orbital $\chi^{n/2}$ of (the neutral) fragment 2, this approximation is small. This results in the improved coupling values obtained above.

In the case of electron transfer, the correct frontier orbitals according to the nomenclature used in eqs.~\ref{eq:simplified_ds_all} and \ref{eq:fo_dft_hab} are $\phi_D^{2n+1}$ and $\phi_A^{2n+1}$, that is, the anionic species of both fragments. For the $H^{\text{2n}}@DA$ scheme this means coupling values are calculated between the LUMOs of both fragments, while in the $2n+1$-schemes the coupling is again calculated between occupied and unoccupied orbitals. For the HAB7 data set the accuracy of the $H^{\text{2n}}@DA$ flavor is best, with an MRSE of $-22.4\,\%$. It is closely followed by the new $H^{\text{2n+1}}@D^-A$ method (MRSE $-22.9\,\%$), while the $H^{\text{2n+1}}@D^-A^-$ variant shows the worst performance (MRSE $-27.1\,\%$). 

Interpretation of these results is much less straightforward than in the cationic case. Due to the known challenges of DFT to correctly describe anionic species\cite{rienstra-kiracofe_atomic_2002,teale_orbital_2008,jensen_describing_2010} one would expect the original $H^{\text{2n}}@DA$ approach -- which does not include any charged calculations by construction -- to perform best, while the $H^{\text{2n+1}}@D^-A^-$ method should perform worst due to being based on two anionic calculations. Following this argument the $H^{\text{2n+1}}@D^-A$ scheme would lie somewhere in the middle. This is at first indeed confirmed by our results. Yet, the fact that the original and our new method here show almost identical performance suggests that this error is at least partially compensated by the less approximate Hamiltonian in the new $H^{\text{2n+1}}@D^-A$ FO-DFT variant.

\subsection{Importance of fragment polarization}
In section~\ref{sec:theory} we explained that by using the superposition of isolated fragments any interactions naturally affecting the molecules in the fragment calculation are neglected. One such interaction is the polarization of a fragment's electron density due the presence of the other. Yet, the severity of this approximation is dependent on the system in question. In Fig.~\ref{fig:hab11_embedding} the comparison between polarized and non-polarized fragment calculations for the HAB11 database showed no significant difference in the estimated couplings (between $-2.3\,\%$ and $+3.2\,\%$). While there certainly is a distortion of the neutral electron density -- as demonstrated in Fig.~\ref{fig:electrostatic_scheme}, neglecting polarization can not alone account for the underestimation of electronic coupling values in the investigated systems. 

\subsection{The influence of exact exchange}\label{sec:disc_hybrids}
When calculating electronic couplings with the B3LYP functional, the MRSE compared to calculations at the BLYP level is reduced from $-22.4\,\%$ to $-7.3\,\%$ ($-22.9\,\%$ to $+11.8\,\%$) for the HAB11 (HAB7) database. This improvement of all electronic couplings, for both hole and electron transfer, computed with hybrid functionals is very remarkable. In particular, since the electronic coupling is proportional to the wave function overlap and the density is more localized (= compact) in hybrid-level calculations, such an increase in coupling values is at first counterintuitive. However, it may again be explained considering the structure of a FO-DFT calculation. Our calculations in Section~\ref{sec:crossover}, using hybrid-level electron densities, but constructing the Hamiltonian with a GGA functional, indeed show the expected (slight) decrease in the coupling values. However, the effect of a Hamiltonian constructed with a hybrid functional based on GGA densities, far outweighs this small reduction due to the more localized electron density. In this case, the electronic couplings are increased by an average of $31.8\pm 8.8\,\%$ ($43.8\pm 12.6\,\%$) compared to the GGA reference. We consistently see this behavior for the B3LYP, PBE0 and HSE06 hybrid functionals, all with similar improvements (see Table~\ref{tab:hybrids}). This effect can thus solely originate in the exact exchange part, since any effects on the charge density are excluded in this approach by using the GGA density.
 
\begin{figure}[ht]
 \centering
 \includegraphics[scale=0.58,clip=true]{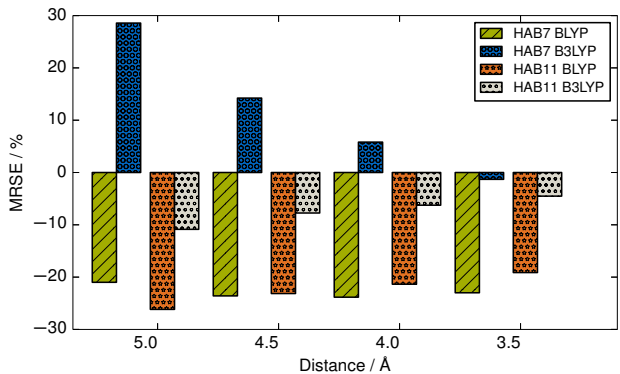}
 \caption{Distance-dependent MRSE for the HAB11 and HAB7 test set. The errors are shown for the BLYP and B3lYP functionals, respectively. The variation in the error is largest for B3LYP in the HAB7 set, while the others change less with the distance.}
 \label{fig:distdep}
\end{figure}

It is important to point out the different performance with respect to anionic and cationic species here. While for hybrid calculations of the cationic species in the HAB11 set both the absolute $H_{ab}$ values and the distance decay behavior $\beta$ are improved, this is not the case for the anionic species in the HAB7 set. The distance decay factor $\beta$ is a sensitive indicator for the distance-dependent error. In the case of B3LYP and HAB7, the accuracy of the electronic couplings varies with the distance, as shown in Fig.~\ref{fig:distdep}. This causes the observed inaccuracy in $\beta$, while still retaining a good overall accuracy on the absolute coupling values. 

\subsection{Computational efficiency versus accuracy}
One of the important advantages of FO-DFT over other methods is its high computational efficiency. Considering a charge transfer dimer of molecules, only the isolated molecules need to be computed in a self consistency cycle, while for the combined (larger) system only a single evaluation of the Kohn-Sham matrix is necessary. Especially in screening studies or disordered systems, where it may be necessary to calculate hundreds or even thousands of electronic coupling values, FO approaches -- sometimes even based on semi-emperical ZINDO\cite{kirkpatrick_approximate_2008, ruhle_microscopic_2011} or DFTB\cite{kubas_electronic_2014,heck_fragment_2014} -- are correspondingly often the method of choice. It is therefore important to point out that the different flavors of FO-DFT not only have different levels of accuracy, but that this accuracy also comes at different computational cost in terms of the number of self consistency cycles necessary (see Table~\ref{tab:acceff}). 

\begin{table}[ht]
  \caption{Compilation of the accuracy of the different FO-DFT flavors in terms of their MRSE for the HAB11 and HAB7 database. The efficiencies of the methods are characterized by the number of full DFT calculations necessary, differentiating between homo- and hetero-dimers. Values for the B3LYP@BLYP crossover scheme are titled B3@B.}
\begin{ruledtabular}
\begin{tabular}{l..ll}
 \multicolumn{1}{l}{  }  & \multicolumn{2}{l}{MRSE / \%} & \multicolumn{2}{l}{$\sum$ calc.} \\ \cline{2-3} \cline{4-5}
 \multicolumn{1}{l}{}  &  \multicolumn{1}{l}{HAB11} & \multicolumn{1}{l}{HAB7} & \multicolumn{1}{l}{homo} & \multicolumn{1}{l}{hetero}\\
  $H^{2n}@DA$ & -24.6 & -22.4  &   1 & 2\\
  $H^{2n\pm 1}@DA/D^-A^-$  & -37.7   & -27.1 &   1 & 2\\
  $H^{2n\pm 1}@D^\pm A$ & -22.4  & -22.9  &   2 & 4\\
  & & & & \\
  $H^{2n\pm 1}@D^\pm A$ (B3LYP) & -7.3  & 11.8  &   2\footnote{full hybrid DFT calculations} & 4\footnotemark[1]\\
  $H^{2n\pm 1}@D^\pm A$ (B3@B) & 2.0  & 11.0  &   2\footnote{only hybrid DFT Hamiltonian in dimer step}  & 4\footnotemark[2]\\
\end{tabular}

\end{ruledtabular}
\label{tab:acceff}
\end{table}

Here it is important to differentiate between homo-dimers (e.g.~in ideal organic crystals) and hetero-dimers (such as in polymers or finite-temperature organic crystals) as well as between hole- and electron-transfer systems. For hole transfer, the best possible accuracy within the FO-DFT approximation is achieved with the charged-fragment scheme ($H^{2n+1}@D^+A$). This comes at the cost of twice as many DFT calculations as for the other schemes. If speed matters most, the original $H^{2n}@DA$ or $H^{2n+1}@DA$ flavors may thus still be more appealing. A word of caution is nevertheless necessary regarding the performance of the $H^{2n}@DA$ method: As emphasized before,\cite{kubas_electronic_2014} the electronic coupling is artificially increased towards the reference value by the spurious excess electron in the Hamiltonian. Since there is no guarantee that this effect does not cause a large overestimation in particular systems, it may be advisable to verify the results against high-level reference data or one of the other FO-DFT schemes.
In the case of electron transfer (i.e.~anionic species), the best accuracy is obtained with the $H^{2n}@DA$ method, closely followed by the new charged-fragment method ($H^{2n+1}@D^-A$). The $H^{2n+1}@D^-A^-$ variant has the worst performance for these systems. 

The introduction of exact exchange via a hybrid DFT functional such as B3LYP yields for both test sets the most accurate electronic couplings attainable with the FO-DFT scheme, yet at a much higher computational cost. A way to reduce this cost and still obtain high-quality electronic couplings may therefore be to use the hybrid-GGA crossover scheme. The most expensive part of the FO-DFT calculation, namely the self consistent calculation of the fragments, is still done on the less demanding GGA level, while only for the final construction of the Hamiltonian the hybrid DFT functional is used.

\section{Conclusions}
In this work we presented a comprehensive evaluation of electronic couplings calculated with different flavors of the efficient fragment orbital scheme. In addition to two previously described variants of FO-DFT we introduced a new scheme resting on slightly different approximations, which lead to an improved description of the diabatic state Hamiltonian. All values were calculated with the same computational settings within the FHI-aims framework, allowing us to rule out any influences of different implementations and technical settings. We compared all calculated values to the high-level \emph{ab initio} references values for the previously introduced HAB7 and HAB11 data sets.\cite{kubas_electronic_2014,kubas_electronic_2015} In accordance with previous work we find that the agreement between values calculated with various DFT frameworks is very good, with differences of typically less than 3\% -- given the same variant of the FO-DFT scheme.

Contrary to earlier expectations\cite{kubas_electronic_2014} we find that hybrid functionals such as PBE0 or B3LYP yield largely improved coupling values for all tested systems. For the new $H^{2n\pm 1}@D^\pm A$ method and B3LYP the MRSE is decreased by $15.1\,\%$ and $11.1\,\%$ for the HAB11 and HAB7 data sets, respectively. This accuracy is then similar to the less approximate constrained DFT approach using hybrid functionals with a tuned exact exchange ratio.\cite{kubas_electronic_2014,kubas_electronic_2015}

We further find that omitting polarization between the fragment densities as common to all fragment orbital schemes has a negligible influence on the electronic coupling value. Overall, the accuracy and performance of the FO-DFT method for systems with weak interactions between the charge transfer sites as often encountered in e.g.~organic semiconductors is thus comparable to more expensive methods such as constrained DFT. This is especially true if the new $H^{2n\pm 1}@D^\pm A$ scheme together with a hybrid DFT functional is used which still shows a more favorable computational cost than a full-fledged hybrid-level constrained DFT calculation.

Based on these results we recommend the following best practice when calculating electronic couplings for hole or electron transfer using FO-DFT: For hole transfer, i.e.~cationic species, the best accuracy is obtained with our new $H^{2n-1}@D^+A$ variant, with an MRSE of only $-22.4\,\%$ for BLYP. If computational efficiency is most important, the classical $H^{2n}@DA$ scheme with an MRSE of $-24.6\,\%$ at GGA-BLYP level performs well. Although it should be noted that part of the improved performance when compared to the similarly effective $H^{2n-1}@DA$ (MRSE $-37.7\,\%$) scheme stems from fortunate error compensation. For electron transfer in anionic species we find a different hierarchy. Here, our new charged fragment scheme has similar accuracy (MRSE $-22.9\,\%$ at GGA-BLYP level) as the original $H^{2n}@DA$ scheme (MRSE $-22.4\,\%$), while the $H^{2n+1}@D^-A^-$ is least accurate (MRSE $-27.1\,\%$). Again, if efficiency is crucial the $H^{2n}@DA$ scheme seems to offer the best compromise between accuracy and efficiency. If the intention is instead to obtain the best possible couplings within the FO-DFT approximation, the $H^{2n\pm 1}@D^\pm A$ method together with a hybrid DFT functional is always the best choice.

\section{Acknowledgements}
The authors gratefully acknowledge support from the Solar Technologies Go Hybrid initiative of the State of Bavaria and the Leibniz Supercomputing Centre for the use of SuperMUC high performance computing facility. We further would like to thank Dr.~Jochen Blumberger for insightful discussions on fragment orbital methods.

\appendix

\section{Applying the Slater-Condon rules}\label{sec:slater}
The step from eq.~\ref{eq:hab} to eq.~\ref{eq:fo_dft_hab} seems to be large, but it can be analytically derived by applying the Slater-Condon rules. The expression for the coupling matrix element $\langle\Psi_a|\hat{\mathcal{H}}|\Psi_b\rangle$,
\begin{align}
H_{ab} = \langle \hat{\mathcal{A}} \phi_D, \ldots, \phi_D^{n-1}, \phi_A, \ldots, \phi_A^{n-1}, \phi_A^n|\mathcal{H}\nonumber\\|\hat{\mathcal{A}} \phi_D, \ldots, \phi_D^{n-1}, \phi_D^n, \phi_A, \ldots, \phi_A^{n-1}\rangle \quad,
\end{align}
with the antisymmetrizing operator $\hat{\mathcal{A}}^2 = \sqrt{N!}\hat{\mathcal{A}}$, can be rewritten as
\begin{align}
H_{ab} = \sqrt{N!}\langle\hat{\mathcal{A}} \phi_D, \ldots, \phi_D^{n-1}, \phi_A, \ldots, \phi_A^{n-1}, \phi_A^n|\mathcal{H}\nonumber\\|\phi_D, \ldots, \phi_D^{n-1}, \phi_D^n, \phi_A, \ldots, \phi_A^{n-1}\rangle \quad.\label{eq:long_full}
\end{align}
If we now replace the exact Hamiltonian with the sum of the Kohn-Sham one-electron Hamiltonians
\begin{align}
\hat{\mathcal{H}}_b = \sum_i^{2n-1}h_b^i \quad,
\end{align}
one can show that of all $N!$ permutations exactly one permutation has a non-zero contribution. 
Only in one case of $\mathcal{P}^{n}$,
\begin{align}
H_{ab} = \langle\phi_D^1, \ldots, \phi_D^{n-1}, {\color{blue}\phi_A^n}, \phi_A^1, \ldots, \phi_A^{n-1}|\mathcal{H}|\nonumber\\\phi_D^1, \ldots, \phi_D^{n-1}, \phi_D^n, \phi_A, \ldots, \phi_A^{n-1}\rangle,\label{eq:long_permuted}
\end{align}
not all summands are zero:

\begin{widetext}
\begin{align}
\underbrace{\langle\phi_D^1|h_b^1|\phi_D^1\rangle}_{=\epsilon^1} \ldots \underbrace{\langle\phi_D^{n-1}|\phi_D^{n-1}\rangle}_{=1} \underbrace{\langle\phi_A^n|\phi_D^n\rangle}_{=0 \text{(orthogonalized!)}}\cdot\underbrace{\langle\phi_A^1|\phi_A^1\rangle}_{=1}\ldots\underbrace{\langle\phi_A^{n-1}|\phi_A^{n-1}\rangle}_{=1} + \nonumber\\
+\underbrace{\langle\phi_D^1|\phi_D^1\rangle}_{1}\ldots\underbrace{\langle\phi_D^{n-1}|h_b^{n-1}|\phi_D^{n-1}\rangle}_{=\epsilon^{n-1}}\cdot \underbrace{\langle\phi_A^n|\phi_D^n\rangle}_{=0} \cdot \underbrace{\langle\phi_A^1|\phi_A^1\rangle}_{=1}\ldots \underbrace{\langle\phi_A^{n-1}|\phi_A^{n-1}\rangle}_{=1} + \nonumber\\
+ \underbrace{\langle\phi_D^1|\phi_D^1\rangle}_{=1}\ldots\underbrace{\langle\phi_D^{n-1}|\phi_D^{n-1}\rangle}_{1}\cdot{\color{blue}\underbrace{\langle\phi_A^n|h_b^n|\phi_D^n\rangle}_{= h_{ab}}}\cdot \underbrace{\langle\phi_A|\phi_A\rangle}_{=1}\ldots \underbrace{\langle\phi_A^{n-1}|\phi_A^{n-1}\rangle}_{=1}\label{eq:right_permutation}
\end{align}
\end{widetext}

This leads to a simplified representation of the diabatic states,
\begin{align}\label{eq:simplified_ds-app}
&\Psi_a \approx \Psi_a^{D^+A} = | \cancel{\hat{\mathcal{A}}\phi_D, \ldots, \phi_D^{n-1}}, {\color{white}\phi_D^n,} \cancel{\phi_A, \ldots, \phi_A^{n-1}}, \phi_A^n\rangle\\
&\Psi_b \approx \Psi_b^{DA^+} = | \cancel{\hat{\mathcal{A}}\phi_D, \ldots, \phi_D^{n-1}}, \phi_D^n, \cancel{\phi_A, \ldots, \phi_A^{n-1}}{\color{white}, \phi_A^n}\rangle
\end{align}%
and the coupling matrix elements
\begin{align}\label{eq:fo_dft_hab-app}
H_{ab} = \langle\phi_A^n|\hat{h}_{b}|\phi_D^n\rangle \quad .
\end{align}

\section{Rotation of wave functions in FHI-aims}\label{sec:localized}
In FHI-aims,\cite{blum_ab_2009} a (numeric atom centered) basis function is defined by
\begin{align}
    \Phi_{i,lm} = \frac{u_i(r)}{r}\cdot S_{l,m}(\theta, \phi),
\end{align}
with a numerically defined function $u_i(r)$ and real-valued spherical harmonics $S_{l,m}(\theta, \phi)$. These are obtained from the complex spherical harmonics $Y_{lm}$ via
\begin{align}
        S_{l,m}(\theta, \phi) = 
    \begin{cases} 
            \frac{(-1)^m}{\sqrt(2)}(Y_{lm} + Y_{lm}^*) & m > 0\\
            Y_{l0} & m=0\\
            \frac{(-1)^m}{i\sqrt(2)}(Y_{l|m|} - Y_{l|m|}^*) & m < 0 \quad,\\
    \end{cases}
\end{align}
even though employing a non-standard sign convention. With the well-known linear combination of atomic orbitals (LCAO) approach,
\begin{align}
        \Psi_k(r) = \sum_{i=1}^{\text{n\_basis}}c_i^k\Phi_i(r) \quad,
\end{align}
one then gets a set of coefficients $c_i^k$ for each wave function. A rotation of a molecule (with a rotation matrix $\mathbf{R}$) leads to the same set of $Y_{lm}$s (as they are fixed with respect to the $xyz$-coordinate system), but with different coefficients $\mathbf{c}$. While different schemes for real spherical harmonics are available (see for example \citet{lessig_efficient_2012,aubert_alternative_2013,blanco_evaluation_1997}), we construct our rotation matrices for different $l$ starting from the complex Wigner~D matrices via a transformation matrix $\mathbf{C}^l$. Using this approach it is easy to account for the different sign convention of FHI-aims in the real spherical harmonics:
\begin{align}
        S_{l,m} = \mathbf{C^l}Y_{l,m} \quad .
\end{align}
The matrix $\mathbf{C}$ is constructed according to \citet{blanco_evaluation_1997} with the constraint of the different sign convention in FHI-aims. With this the real rotation matrix $\Delta^l(R)$ is calculated:
\begin{align}
        \Delta^l(R) = \left(\mathbf{C}^l\right)^*\mathbf{D}^l(R)\left(\mathbf{C}^l\right)^t \quad .
\end{align}

\bibliography{clean_library}
\end{document}